\documentclass[journal]{IEEEtran}

\ifCLASSINFOpdf
\else
   \usepackage[dvips]{graphicx}
\fi
\usepackage{amsmath,graphicx}
\usepackage{booktabs, makecell}
\usepackage{float}
\usepackage{multicol}
\usepackage{subcaption}
\usepackage{amsfonts}
\usepackage{mathtools}
\usepackage{multirow}
\usepackage{bm}
\usepackage{nicefrac}
\usepackage{tabularx}
\usepackage{xcolor}
\usepackage{amsthm}
\usepackage{extarrows}
\usepackage{balance}
\usepackage[noend]{algpseudocode}
\usepackage{algorithm}
\usepackage{url}

\newcommand{\eref}[1]{(\ref{#1})}

\newcommand{\fref}[1]{Fig.~\ref{#1}}
\newcommand{\tref}[1]{Table~\ref{#1}}

\renewcommand{\arraystretch}{1.1}
\newcommand{\ra}[1]{\renewcommand{\arraystretch}{#1}}

\newcommand*\Let[2]{\State #1 $\gets$ #2}
\algrenewcommand\algorithmicrequire{\textbf{Precondition:}}
\algrenewcommand\algorithmicensure{\textbf{Postcondition:}}

\def\bphi{\bm{\phi}}
\def\btheta{\bm{\theta}}

\begin{document}

\title{Scaling Up Adaptive Filter Optimizers}

\author{Jonah Casebeer,~\IEEEmembership{Student Member, IEEE}, Nicholas J. Bryan,~\IEEEmembership{Member, IEEE}, Paris Smaragdis,~\IEEEmembership{Fellow, IEEE}%
\thanks{J. Casebeer is with the Department of Computer Science, University of Illinois at Urbana-Champaign, Urbana, IL 61801 USA~(e-mail: jonah.casebeer@ieee.org).\\
N. J. Bryan is with Adobe Research, San Francisco, CA, 94103 USA~(e-mail: njb@ieee.org)\\
P. Smaragdis is with the Department of Computer Science and Department, University of Illinois at Urbana-Champaign, Urbana, IL 61801 USA~(e-mail: paris@illinois.edu)}
}

\markboth{Journal of \LaTeX\ Class Files, Vol. 14, No. 8, August 2015}
{Shell \MakeLowercase{\textit{et al.}}: Bare Demo of IEEEtran.cls for IEEE Journals}
\maketitle

\begin{abstract}
We introduce a new online adaptive filtering method called supervised multi-step adaptive filters~(SMS-AF). Our method uses neural networks to control or optimize linear multi-delay or multi-channel frequency-domain filters and can flexibly scale-up performance at the cost of increased compute -- a property rarely addressed in the AF literature, but critical for many applications. To do so, we extend recent work with a set of improvements including feature pruning, a supervised loss, and multiple optimization steps per time-frame. These improvements work in a cohesive manner to unlock scaling. Furthermore, we show how our method relates to Kalman filtering and meta-adaptive filtering, making it seamlessly applicable to a diverse set of AF tasks. We evaluate our method on acoustic echo cancellation~(AEC) and multi-channel speech enhancement tasks and compare against several baselines on standard synthetic and real-world datasets. Results show our method performance scales with inference cost and model capacity, yields multi-dB performance gains for both tasks, and is real-time capable on a single CPU core.
\end{abstract}


\IEEEpeerreviewmaketitle

\section{Introduction}
\IEEEPARstart{A}{daptive} filters~(AF) play an indispensable role in a wide array of signal processing applications such as acoustic echo cancellation, equalization, and interference suppression.
AFs are parameterized by time-varying filter weights and require an update or optimization rule to control them over time.
Improving the performance of AFs continues to pose an intricate challenge, requiring a nuanced approach to optimizer design.
Consequently, AF algorithm designers have relied on mathematical insights to create tailored optimizers, starting from the foundational development of the least mean squares algorithm (LMS)~\cite{widrow1960adaptive} to the Kalman filter~\cite{mathews1991adaptive, apolinario2009qrd, haykin2008adaptive, rabiner2016theory}.
\begin{figure}[!tb]
	\centering
	\includegraphics[width=.98\linewidth]{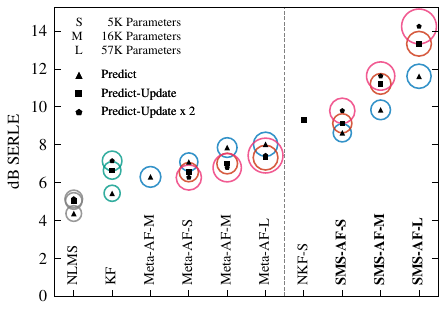}
	\caption{Acoustic echo cancellation performance vs. model size, optimization steps per time-frame~(opt. steps), and supervision levels. Bubble size shows real-time-factor~(RTF) where smaller is faster, inner-shape shows opt. steps, and the vertical dotted line separates unsupervised~(left) and supervised~(right) approaches. SMS-AF, in bold on the far right, demonstrate robust scaling performance in terms of parameters, and RTF. 
 }
	\label{fig:aec_scaling}
\end{figure}

In contrast, we have witnessed countless remarkable deep learning algorithm advancements in other domains through the principle of ``scaling"~\cite{caballero2022broken, hoffmann2022training, liu2022convnet, kang2023scaling}. The scaling approach involves improving an existing method by deploying additional computational resources. Scaling methodologies are particularly enticing, as they tap into the increasing computational capabilities of modern smart devices, minimizing the need for labor-intensive manual tuning and intervention.
In the context of neural networks for online low-latency AFs that can benefit from scaling, we find two general approaches: 1) model-based methods that integrate deep neural networks~(DNNs) into existing AF frameworks to update optimizer statistics~\cite{casebeer2021nice, haubner2022end, haubner2022deep}, step-size~\cite{ivry2022deep, soleimani2023neural}, or other quantities~\cite{yang2023low} and 2) model-free strategies that do not rely on an existing AF strategy, learn AF optimizers using meta-learning in an end-to-end fashion~\cite{casebeer2021auto, casebeer2022meta, wu2022meta, casebeer2023metaaf}, and yield state-of-the-art (SOTA) results~\cite{yang2023low, casebeer2023metaaf}.
We focus on the latter, given their recent success, but note scaling such approaches have either been limited to high-latency regimes~\cite{wu2022meta} or only marginally improves results~\cite{yang2023low}.

We propose a new online AF method called supervised multi-step AF (SMS-AF). Our method integrates a series of algorithm improvements on top of recent meta-learning methods~\cite{casebeer2022meta, wu2022meta} that together enable scaling performance by increasing model capacity and/or inference cost as shown in~\fref{fig:aec_scaling}.
We evaluate our approach on the tasks of acoustic echo cancellation~(AEC) and generalized sidelobe canceller~(GSC) speech enhancement and compare to recent SOTA approaches. Results show that our scaling behavior translates to substantial performance gains in all metrics across tasks and datasets and delivers breakthrough AEC performance

Our contributions include:
1) A new general purpose AF method that allow us to reliably improve performance by simply using more computation,
2) Design insights for customizing our proposed method for the task of AEC and GSC,
3) Empirical exploration of AF optimizer scaling showing our approach scales vs. model size and optimizer step count, and
4) Insights as to how our approach generalizes Kalman filtering.

\def\time{{\mathrm{t}}}
\def\freq{\mathrm{k}}
\def\mic{{\mathrm{m}}}
\def\buffer{\mathrm{b}}
\def\frame{\tau}

\def\F{\mathbf{F}}
\def\I{\mathbf{I}}
\def\0{\mathbf{0}}
\def\1{\mathbf{1}}

\def\u{\mathbf{u}}
\def\U{\mathbf{U}}
\def\d{\mathbf{d}}
\def\D{\mathbf{D}}
\def\y{\mathbf{y}}
\def\e{\mathbf{e}}
\def\s{\mathbf{s}}
\def\w{\mathbf{w}}

\def\H{\mathsf{H}}
\def\T{\top}
\def\diag{\operatorname{diag}}

\def\L{\mathcal{L}}
\def\grad{\bm{\nabla}}

\section{Background}

\subsection{Adaptive Filters}
\label{sec:background}
An AF is an optimization procedure that seeks to adapt filter parameters to fit an objective over time. AFs typically input a mixture $\underline{\d}[\frame]$, adjust a time-varying linear filter $h_{\btheta[\frame]}$ with parameters $\btheta[\frame]$ to remove noise via knowledge from a reference signal $\underline{\u}[\frame]$, produce estimate $\underline{\y}[\frame]$, and output an error signal $\underline{\e}[\frame]$. We focus on  multi-delay and/or multi-channel frequency-domain filters (MDF) for low-latency processing.
The filter parameters are updated across time by minimizing a loss, $\L(h_{\btheta[\frame - 1]},\cdots)$, resulting in a per frame $\frame$ update rule,
\begin{equation}
    \btheta[\frame] = \btheta[\frame - 1] + \bm{\Delta}[\frame].
\label{eq:general_update}
\end{equation}
$\bm{\Delta}[\frame]$ can also be written as the output of an optimizer $g_{\bphi}(\bm{\xi}[\frame]) = \bm{\Delta}[\frame]$ with input $\bm{\xi}[\frame]$, and parameters $\bphi$.

\subsection{Adaptive Filter Optimizers}
The AF optimizer, $g_{\bphi}(\bm{\xi}[\frame])$ is key and can vary in levels of sophistication. In the simple case, the optimizer can be a hand-derived algorithm such as LMS. In this case, each parameter in $\btheta$ is updated independently, so $g$ accepts the gradient with respect to the loss via $\bm{\xi}[\frame]$, and is only parameterized by the step-size $\bphi=\{\lambda\}$, resulting in $g$ simply scaling the gradient.
Most AFs operate via the following steps~\cite{haykin2008adaptive}:
\begin{eqnarray}
    \underline{\e}[\frame] &=& h_{\btheta[\frame - 1]}(\underline{\d}[\frame], \underline{\u}[\frame])\label{eq:predict}\\
    \bm{\Delta}[\frame] &=& g_{\bphi}(\bm{\xi}[\frame], \cdots) \label{eq:opt_forward}\\
    \btheta[\frame] &=& \btheta[\frame - 1] + \bm{\Delta}[\frame]\label{eq:fil_update},
\end{eqnarray}
where~\eref{eq:predict} applies the filter,~\eref{eq:opt_forward} updates the optimizer, and~\eref{eq:fil_update} updates the filter parameters for the next frame. Optimizers typically use filter output $\underline{\e}[\frame]$, produced via $\btheta[\frame-1]$, creating a feedback loop.
The Kalman filter (KF) extends the above via distinct ``predict" and ``update" steps. In the KF predict step, \eref{eq:predict}-\eref{eq:fil_update} are run as normal. In the KF update step, however, the filter output is reprocessed after \eref{eq:fil_update} using the latest data:
\begin{eqnarray}
    \underline{\e}[\frame] &=& h_{\btheta[\frame]}(\underline{\d}[\frame],\underline{\u}[\frame])\label{eq:predict2}.
\end{eqnarray}

\subsection{Learned Optimizers}
Historically, AFs are hand-derived, given a loss and filter. In contrast, neural-AF optimizers can be trained via meta-learning (Meta-AF)~\cite{andrychowicz2016learning, casebeer2021auto, casebeer2022meta}. Meta-AFs are trained to control MDF filters via recurrent neural networks (RNNs) with parameters $\bphi$ that are trained to maximize AF performance on a large dataset via an unsupervised (or self-supervised) meta-objective $\L_M(\; g_{\bphi}, \L(h_{\btheta}, \cdots) \;)$ and backpropagation through time (BPTT).
A common meta-loss is,
\begin{eqnarray}
    \L_M(\underline{\bar{\e}}) &=& \ln E[\|\underline{\bar{\e}}[\frame]\|^2],
     \label{eq:metaloss}
\end{eqnarray}
where $\underline{\bar{\e}}[\frame] = \mathrm{cat}(\underline\e[\frame], \cdots, \underline\e[\frame+L-1]) \in \mathbb{R}^{RL}$, $L$ is the truncation length, $R$ is the hop size, and $\mathrm{cat}$ concatenates.

Two important extensions to Meta-AF include 1) higher-order Meta-AF (HO-Meta-AF)~\cite{wu2022meta}, which introduces learnable coupling modules to model groups of filter parameters, reduce complexity, and improve performance for high-latency single-block frequency-domain filters and 2) low-complexity neural Kalman filtering (NKF)~\cite{yang2023low}, which extends Meta-AF with a KF, a supervised loss, and different training setup. We regard Meta-AF as SOTA for unsupervised AFs (see Table IV~\cite{casebeer2022meta}) and NKF as SOTA for supervised AFs~\cite{yang2023low}.

\section{Scaling Up Learned Optimizers}
As the foundation of our SMS-AF method, we combine Meta-AFs~\cite{casebeer2022meta} with a higher-order optimizer~\cite{wu2022meta} with per-frequency inputs $\bm{\xi}_{\freq}[\frame]$, and then extend it with three task-agnostic improvements and one task-specific change. Our training and inference methods are summarized in Alg.~\ref{alg:training}.

\subsection{Scaling Up Feature Quality: Feature Pruning}\label{label:feature}
Our first insight is to use only three features to control filter adaptation: knowledge of the filter input, final filter output, and filter state.
Compared to past work~\cite{casebeer2022meta} that uses,
\begin{eqnarray}
\bm{\xi}_{\freq}[\frame] = [\nabla_{\freq}[\frame], \u_{\freq}[\frame], \d_{\freq}[\frame], \e_{\freq}[\frame], \y_{\freq}[\frame]],
\end{eqnarray}
where $\nabla_{\freq}[\frame]$ is a gradient w.r.t. loss $\L(\cdots)$, we use
\begin{eqnarray}
\bm{\xi}_{\freq}[\frame] = [\u_{\freq}[\frame], \e_{\freq}[\frame], \btheta_{\freq}[\frame]].
\end{eqnarray}
Pruning reduces complexity by lowering input dimension and memory requirements for the optimizer,
while eliminating inference-time gradients, as shown in line $8$ of Alg.~\ref{alg:training}.
\subsection{Scaling Up Supervision: Supervised Loss}\label{label:loss}
Our second insight is to use a high-quality supervised loss, instead of an unsupervised loss.
Previous methods have explored supervised losses such as frame-wise independent supervised losses for echoes~\cite{yang2023low} or oracle filter parameters~\cite{haubner2022end}. These methods treat frequency bins, adjacent frames, and other channels as distinct optimization entities and have not scaled~\cite{yang2023low}. As such, we compute our supervised loss in the time-domain after all AF operations have been performed. This strategy is similar to~\eref{eq:metaloss}, but with supervision. Our supervision is non-causal; the loss at $\frame$ depends on updates from $<\frame$, enabling the optimizer to learn anticipatory updates.
For AEC,
\begin{eqnarray}
    \L_S(\d_{\u}, \e) &=& \ln E[\|\underline{\bar{\d}}_{\u}[\frame] - \underline{\bar{\e}}[\frame]\|^2],
     \label{eq:supmetaloss}
\end{eqnarray}
where $\d_{\u}[\frame] = \u[\frame] \ast \w[\frame]$ is the true echo. For GSC, we use scale-invariant signal-to-distortion ratio~(SI-SDR).
Better loss functions exclusively impact the training phase, without contributing to test-time complexity, making this change cost-free for inference. This corresponds to line $22$ of Alg.~\ref{alg:training}.

\subsection{Scaling Up Feedback: Multi-Step Optimization}\label{label:multi}
Our third insight is to leverage the iterative nature of optimizers by executing multiple optimization steps per time frame. By doing so, we offer our optimizers a more powerful feedback mechanism and use the most current parameters for the filter output.
Specifically, we run our optimizer update via
\eref{eq:predict}-\eref{eq:fil_update},  \eref{eq:predict}-\eref{eq:predict2}, or looping over \eref{eq:predict}-\eref{eq:predict2} multiple times. The first option follows Meta-AF, the second option follows a typical KF, and the third extends the KF. We denote the number of \eref{eq:predict}-\eref{eq:fil_update} iterations via $C$.
Incorporating multi-step optimization in Alg.~\ref{alg:training} involves three changes. First, initializing each frame's filter and optimizer state with results from the last frame~(lines $3-4$). Second, iteratively progressing through steps withing a frame~(line $5$), while updating filter parameters/outputs, and optimizer state~(lines $8-10$). Last, running a final filter forward pass using the latest parameters~(line $11$).
We find this approach to be a compelling alternative to increasing the dimension of the optimizer, $H$. Notably, it avoids increasing the parameter count, and it linearly scales complexity, in stark contrast to the quadratic complexity effects associated with $H$.

\begin{algorithm}[t]
  \caption{Training and inference algorithm.}
  \begin{algorithmic}[1]
    \Function{Forward}{$g_{\bphi}, h_{\btheta}, \underline\u, \underline\d_\mic$}
      \For{$\frame \gets 0 \textrm{ to } L $}

        \Let{$\btheta_{\freq}[\frame]$}{$\btheta_{\freq}[\frame - 1]$}
        \Comment{Initialize with last estimate}
        \Let{$\boldsymbol{\psi}_{\freq}[\frame]$}{$\boldsymbol{\psi}_{\freq}[\frame - 1]$}
        \Comment{Initialize with last state}

        \For{$\mathrm{c} \gets 0 \textrm{ to } C$}
        \Comment{For each PU iteration}
            \Let{$\underline{\e}[\frame]$}{$h_{\btheta[\frame]}(\underline{\d}[\frame], \underline{\u}[\frame])$}

            \For{$\freq \gets 0 \textrm{ to } K$}
            \Comment{Predict step}
              \Let{$\boldsymbol{\xi}_{\freq}[\frame]$}{$[\u_{\freq}[\frame], \e_{\freq}[\frame], \btheta_{\freq}[\frame]$}

              \Let{$(\boldsymbol{\Delta}_{\freq}[\frame], \boldsymbol{\psi}_{\freq}[\frame])$}{$g_{\bphi}(\boldsymbol{\xi}_{\freq}[\frame], \boldsymbol{\psi}_{\freq}[\frame])$}

              \Let{$\btheta_{\freq}[\frame]$}{$\btheta_{\freq}[\frame] +  \boldsymbol{\Delta}_{\freq}[\frame]$}
            \EndFor
        \EndFor

        \Let{$\underline{\e}[\frame]$}{$h_{\btheta[\frame]}(\underline{\d}[\frame], \underline{\u}[\frame])$}
        \Comment{Update step}
      \EndFor
         \Let{$\bar{\e}$}{\textproc{Cat}$(\underline\e[\frame] \; \forall \frame)$}
      \State \Return{$\bar{\e}, \boldsymbol{\psi}[\frame], h_{\btheta[\frame]}$}
    \EndFunction

    \Function{Train}{$\mathcal{D}$}
      \Let{$\bphi$}{\cite{wolter2018complex} init}
      \While{$\bphi$ \textbf{not} \textproc{Converged}}
          \Let{$\underline\u, \underline\d_\mic$}{\textproc{Sample}($\mathcal{D}$)}
          \Comment{Get batch from $\mathcal{D}$}
          \Let{$\btheta, \boldsymbol{\psi}$}{$\mathbf{0}, \mathbf{0}$}
          \For{$n \gets 0 \textrm{ to end }$}
          \Let{$\underline\u, \underline\d_\mic$}{\textproc{NextL}$(\underline\u, \underline\d_\mic)$}          \Comment{Grab $L$ frames}
          \Let{$\underline{\u}, \boldsymbol{\psi}, h_{\btheta}$}{\textproc{Forward}($g_{\bphi}, \boldsymbol{\psi}, h_{\btheta}, \underline\u, \underline\d_\mic$)}

          \Let{$L_S$}{$\L_S(\cdots)$}
          \Comment{Task-dependent objective}

          \Let{$\grad$}{\textproc{Grad}($L_S$, $\bphi$)}
          \Comment{Truncated BPTT}
          \Let{$\bphi[n + 1]$}{\textproc{MetaOpt}($\bphi[n], \grad$)}
          \Comment{i.e. Adam}
          \EndFor
      \EndWhile
      \Return{$\hat{\bphi}$}
    \EndFunction
  \end{algorithmic}
  \label{alg:training}
\end{algorithm}

\subsection{Modifications to Overlap-Save for AEC}\label{label:os}
When using overlap-save, we noticed artifacts due to rapid filter adaptation. So, we applied a straightforward solution: overlap-add with a synthesis window, but no analysis window.
\subsection{Perspectives}
Our modifications are a notable departure from the Meta-AF methodology, but still aim to learn a neural optimizer for AFs end-to-end.
First, by pruning inputs features and introducing a supervised loss, we eliminate the need for explicit meta-learning, leading to a more streamlined BPTT training process.
Second, we replace the past unsupervised loss with a new, strong supervised signal and loss, helping us scale up.
Third, we leverage a multi-step optimization scheme. This creates a generalization of the Kalman filter, where all parameters are entirely learned, while retaining explicit predict and update steps. This also effectively deepens our optimizer networks by sharing parameters across layers in a depth-wise manner.

\section{Experimental Design}

\subsection{Experiments}
The goal of our experiments is to benchmark SMS-AF and demonstrate how it scales. To do so, we perform an initial within method ablation, then study scaling on AEC and GSC tasks and vary 1) optimizer model sizes with small~(S), medium~(M), and large~(L) models 2) an unsupervised~(U) or supervised~(S) loss and 3) the number of predict~(P) and update~(U) steps per frame. Each experiment is labeled with an identifier (e.g. S$\cdot$S$\cdot$PU), indicating the size, supervision, and number of PU steps. We label baselines when applicable.

\subsection{AEC Configuration}
Our AEC signal model is $\underline{\d}[\time] =\underline{\u}[\time]\ast\underline{\w}+\underline{\mathbf{n}}[\time]+\underline{\mathbf{s}}[\time]$, where $\underline{\mathbf{n}}$ stands for noise, and $\underline{\mathbf{s}}$ is speech. The goal is to recover the speech $\underline{\mathbf{s}}$ given the far-end $\underline{\u}$, and mixture $\underline{\d}$. This involves fitting a filter to mimic $\underline{\w}$. We use a linear MDF filter with $8$ blocks, each of size $512$, a hop of $256$, and construct the output using overlap-add with a Hann window.
Our baselines are NLMS, KF~\cite{enzner2006frequency}, Meta-AF~\cite{casebeer2022meta}, HO-Meta-AF~\cite{wu2022meta}, and Neural-Kalman Filter~\cite{yang2023low}. We also test several HO-Meta-AF model sizes as well as multi-step NLMS, KF, and HO-Meta-AF. For training, we use the synthetic fold of the Microsoft AEC Challenge~\cite{cutler2022AEC}.
Each scene has double-talk, near-end noise and loud-speaker nonlinearities. We also evaluate on the real, crowd-sourced, blind test-set~\cite{cutler2022AEC}.
We use echo return loss enhancement~(ERLE)~\cite{enzner2014acoustic} to measure echo reduction. To describe perceptual quality, we use AEC-MOS, a reference-free model that predicts a 5 point score~\cite{cutler2022AEC}. On real data, we prefix with an R, use ERLE in single-talk and AEC-MOS in double-talk. To quantify complexity, we use mega FLOP~(MFLOP) counts, single core real-time-factor~(RTF) equal to processing over elapsed time, and model size.
\subsection{GSC Configuration}
For GSC, we use a single-block frequency-domain GSC beamformer.
The signal model at each of the $M$ microphones is $ \underline{\u}_{\mic}[\time] = \underline{\mathbf{r}}_{\mic}[\time] \ast \underline{\s}[\time] + \underline{\mathbf{n}}_{\mic}[\time]$, and $\underline{\mathbf{r}}_{\mic}[\time]$ is the impulse response from source to mic $\mic$. The goal is to recover the clean speech $\underline{\s}$ given the input signal $\underline{\u}_{\mic}[\time]$. This requires fitting a filter to remove the effects of noise, $\underline{\mathbf{n}}_{\mic}$. We assume access to a steering vector
and compare against NLMS, recursive-lease-squares~(RLS), and Meta-AF. We test multiple HO-Meta-AF model sizes as well as multi-step NLMS, RLS, and HO-Meta-AF. We used the CHIME-3~\cite{barker2015third} dataset.
For overall quality, we compute scale-invariant signal-to-distortion ratio~(SI-SDR)~\cite{le2019sdr}, and contrast
signal-to-interference ratio (SIR), and signal-to-artifact ratio~(SAR)~\cite{vincent2006performance}. For perceptual quality, we use Short-Time Objective Intelligibility (STOI)~\cite{taal2011algorithm}.
\subsection{Optimizer Configuration}
For AEC and GSC, we use higher-order Meta-AF optimizers with banded coupling, and a group size of $5$~\cite{wu2022meta}.
This amounts to a Conv1D layer, two GRU layers, and a transposed Conv1D layer.
To train, we use Adam with a batch of $16$, a learning rate of $10^{-4}$, and randomize the truncation length $L$ with a maximum of $128$. We apply gradient clipping and reduce the learning rate by half if the validation performance does not improve for $10$ epochs, and stop training after $30$ epochs with no improvement. We use log-MSE loss on the echo for AEC, and SI-SDR loss on the clean speech for GSC. All models are trained on one GPU. Note, our S, M, and L model sizes correspond to $g_{\bphi}$ hidden state sizes of $16$, $32$, and $64$ with parameters counts of about $5$K, $16$K, and $57$K.

\section{Results \& Discussion}
\setlength{\tabcolsep}{1pt}
\begin{table}[!t]
    \centering
    \ra{.5}
    \caption{AEC performance vs. computational cost.}
    \vspace{-1mm}
    \begin{tabular*}{1\linewidth}{@{\extracolsep{\fill}}l c r r r r r r r r r r r r r@{}}\toprule
        Model && ERLE$\uparrow$ && \makecell{R-ERLE$\uparrow$} &&  \makecell{AEC-\\MOS$\uparrow$} && \makecell{R-AEC-\\MOS$\uparrow$} && MFLOPs$\downarrow$ && RTF$\downarrow$\\ \midrule
        Mixture && - &&  - && 2.33 && 2.69 && - && -\\
        NLMS$\cdot$P && 4.37 &&  1.74 && 3.16 && 3.06 && 0.08 && 0.31\\
        KF$\cdot$P &&  5.44 && 2.71 && 3.32 &&  3.18 && 0.12 && 0.32\\
        KF$\cdot$PU &&  6.65 && 3.83 && 3.65 && 3.38 &&  0.12 && 0.35\\
        KF$\cdot$PUx2 && 7.15 && 5.58 && 3.89 && 3.66 && 0.18 && 0.39\\
        \midrule
        M$\cdot$U$\cdot$P~\cite{casebeer2022meta} && 6.30 && 2.99 && 3.62 &&3.31 && 9.02 && 0.41\\
        S$\cdot$U$\cdot$P && 7.09 && 3.30 && 3.57 && 3.26 && 2.80 && 0.36\\
        M$\cdot$U$\cdot$P && 7.85 && 3.75 && 3.62 && 3.30 && 7.07 && 0.39\\
        L$\cdot$U$\cdot$P && 8.03 && 3.91 && 3.65 && 3.33 && 20.42 && 0.48\\
        \midrule
        NKF$\cdot$S$\cdot$PU &&  9.29 &&  6.38 && 3.69 && 3.59 && 10.16 && -\\
        S$\cdot$S$\cdot$P && 8.63 && 3.81 && 3.63 && 3.31 && 2.80 && 0.36\\
        M$\cdot$S$\cdot$P && 9.84 && 4.62 && 3.73 && 3.40 && 7.07 && 0.39\\
        L$\cdot$S$\cdot$P && 11.62 && 5.52 && 3.83 && 3.50 && 20.42 && 0.48\\
        \midrule
        S$\cdot$S$\cdot$PU && 9.13 && 6.05 && 3.85 && 3.55 && 2.81 && 0.38\\
        M$\cdot$S$\cdot$PU && 11.22 && 7.87 && 3.99 && 3.72 && 7.08 && 0.41 \\
        L$\cdot$S$\cdot$PU && 13.34 && 9.78 && 4.05 && 3.85 && 20.43 && 0.49\\
        \midrule
        S$\cdot$S$\cdot$PUx2 && 9.80 && 6.96 && 3.85 && 3.66 && 5.56 && 0.50\\
        M$\cdot$S$\cdot$PUx2 && 11.77 &&  8.86 && 4.04 && 3.80 && 14.11 && 0.56\\
        L$\cdot$S$\cdot$PUx2 && \textbf{14.25} && \textbf{11.15} && \textbf{4.12} && \textbf{3.94} && 40.81 && 0.69\\
        \bottomrule
    \end{tabular*}
    \label{tab:aec_leaderboard}
    \vspace{-2mm}
\end{table}

\subsection{Initial Within Method Ablation}
We perform an initial within-method ablation on the task of AEC to understand our modifications.
First, we compare an M$\cdot$U$\cdot$P variant trained with the full feature vs. pruned feature set. The full feature set achieves an ERLE of $6.39$dB~(not shown), while the pruned set scores $7.85$dB, over a dB gain.
We expand on the pruned model and add supervision, resulting in an ERLE improvement to $9.84$dB, a gain of nearly $2$dB.
We then use multiple steps per frame. Extending the supervised model with an update-step increases ERLE to $11.22$dB, and doubling the iterations reaches $11.77$dB.
To mitigate clicking artifacts, we then add our modified OLA scheme. This change reduces the ERLE by $\approx1$ dB ERLE, but removes severe clicking artifacts.
Combined, this yields a $4$dB ERLE gain.

\subsection{AEC Scaling Ablation and Benchmarking}
Next, we explore scaling in AEC as shown in~\fref{fig:aec_scaling} and~\tref{tab:aec_leaderboard}. We attempt to scale up our baselines and then do so with our proposed model.
When scaling model size, we notice that scaling the unsupervised model from S to L (S$\cdot$U$\cdot$P to L$\cdot$U$\cdot$P) results in a peak gain of $\approx 1$dB, to $8.03$dB ERLE.
In contrast, scaling the supervised model from S to L (S$\cdot$S$\cdot$P to L$\cdot$S$\cdot$P) yields larger gains, peaking at $11.62$dB.
When scaling optimization steps, we find the unsupervised models from S$\cdot$U$\cdot$P to S$\cdot$U$\cdot$PUx2 results in marginal or even a negative performances changes.
In contrast, scaling from S$\cdot$S$\cdot$P to S$\cdot$S$\cdot$PUx2 provides $+1$dB, and L$\cdot$S$\cdot$P to L$\cdot$S$\cdot$PUx2 provides $+2.65$dB, showing supervision is crucial to unlock the benefit of multiple opt. steps per frame.
Our best$\cdot$performing L$\cdot$S$\cdot$PUx2 scores over $14$dB ERLE, doubling the S$\cdot$U$\cdot$P performance of $7.09$dB.

When benchmarking against competing methods, we note that SOTA sueprvised NKF method is most comparable. Our S$\cdot$S$\cdot$PU model matches NKF performance while using only one-fifth of the NKF MFLOP count. Our M$\cdot$S$\cdot$PU model further enhances all metrics and uses fewer MFLOPs. For our best-performing L$\cdot$S$\cdot$PUx2, we score $14.25$dB ERLE, a $4.96$dB improvement over NKF.
In perceptual metrics, our top-performing L$\cdot$S$\cdot$PUx2 model achieves $11$dB in R-ERLE and a R-AEC-MOS of $3.94$, while remaining real-time on a single CPU core.
Surprisingly, RTF scales non-linearly with MFLOPs and model size, showing untapped scaling potential.

\setlength{\tabcolsep}{1.15pt}
\begin{table}[!t]
    \centering
    \ra{.5}
    \caption{Beaforming performance vs. computational cost.}
    \vspace{-1mm}
    \begin{tabular*}{.99\linewidth}{@{\extracolsep{\fill}}l r r r r r r r r r r r r@{}}\toprule
        Model && SI-SDR$\uparrow$ && SIR$\uparrow$ && SAR$\uparrow$ && STOI$\uparrow$ && MFLOPs$\downarrow$ && RTF$\downarrow$\\ \midrule
        Mixture && -0.71 && - && - && 0.674 && - && - \\
        NLMS$\cdot$P && 8.60 && 16.21 && 9.78 && 0.905 && 0.43 && 0.36 \\
        NLMS$\cdot$PU && 8.84 && 16.54 && 10.00 && 0.910 && 0.47 && 0.47\\
        RLS$\cdot$P && 9.84 && 16.70 && 9.70 && 0.919 && 0.53 && 0.50\\
        RLS$\cdot$PU && 10.14 && 17.16 && 11.49 && 0.924 && 0.54 && 0.62\\
        \midrule
        S$\cdot$U$\cdot$P && 12.20 && 22.57 && 12.79 && 0.931 && 4.70 && 0.41\\
        M$\cdot$U$\cdot$P && 12.62 && 22.56 && 13.26 && 0.938 && 12.08 && 0.45\\
        L$\cdot$U$\cdot$P && 12.45 && 22.43 && 13.09 && 0.935 && 36.35 && 0.53\\
        \midrule
        S$\cdot$S$\cdot$P && 13.92 && 23.00 && 14.66 && 0.950 && 4.70 && 0.41\\
        M$\cdot$S$\cdot$P && 14.34 && 23.45 && 15.07 && 0.953 &&  12.08 && 0.45\\
        L$\cdot$S$\cdot$P && 14.69 && 24.36 && 15.33 && 0.954 && 36.35 && 0.53\\
        \midrule
        S$\cdot$S$\cdot$PU && 15.46 && 25.69 && 16.09 && 0.956 && 4.74 && 0.51\\
        M$\cdot$S$\cdot$PU && 16.83 && 27.70 && 17.41 && 0.960 && 12.12 && 0.54 \\
        L$\cdot$S$\cdot$PU && 17.22 && 28.37 && 17.80 && 0.962 && 36.39 && 0.62\\
        \midrule
        S$\cdot$S$\cdot$PUx2 && 15.67 && 25.89 && 16.35 && 0.956 && 9.07 && 0.70\\
        M$\cdot$S$\cdot$PUx2 && 17.06 && 28.42 && 17.63 && 0.961 && 23.83 && 0.76\\
        L$\cdot$S$\cdot$PUx2 && \textbf{17.72} && \textbf{29.52} && \textbf{18.25} && \textbf{0.964} && 72.37 && 0.91\\
        \bottomrule
    \end{tabular*}
    \label{tab:gsc_leaderboard_all}
    \vspace{-2mm}
\end{table}

\subsection{GSC Beamforming Scaling and Benchmarking}
Beamforming results are in~\tref{tab:gsc_leaderboard_all}. Notably, SMS-AF improvements apply without any modifications. Here, all models assume access to a steering vector, which can be challenging to estimate in practice. Again, supervision and multi-step optimization yield significant performance gains. Our S$\cdot$S$\cdot$P model outperforms all baselines including L$\cdot$U$\cdot$P across all metrics. Our model scales reliably with the L$\cdot$S$\cdot$P variant improving performance in all metrics. Scaling up the iterations to S$\cdot$U$\cdot$PUx2 yields larger gains across all metrics. Our largest and best model, L$\cdot$S$\cdot$PUx2 scores a remarkable 17.72 dB SI-SDR while still being real-time. Of note, the L$\cdot$S$\cdot$PU model has the same RTF as RLS$\cdot$PU, even though RLS uses fewer operations.
Again, we show that SMS-AF performance scales with both model capacity and optimization steps per frame.

\section{Conclusion}
We introduce a method for a neural network-based adaptive filter optimizers called supervised multi-step adaptive filters (SMS-AF). We extend meta-adaptive filtering methods with several advances that combine to reliably increase performance by leveraging more computation.
We evaluate our method on low latency, online AEC and GSC tasks, compare against many baselines and test on both synthetic and real data.
SMS-AF improves both subjective and objective metrics, achieving $\approx 5$dB ERLE/SI-SDR gains compared to prior work, and increases the performance ceiling across AEC and GSC.
Furthermore, we relate our work to the Kalman filter and meta-AFs, giving insight for many other applications.
We believe scaling-up AFs is a promising direction and hope our results encourage future work on scalable, general purpose AFs.

\bibliographystyle{IEEEbib}
\bibliography{strings,refs}

\end{document}